\documentstyle[aps,preprint,epsfig,floats]{revtex}

\tightenlines

\begin{document}
\title{Formation of Structure in Snowfields: \\
Penitentes, Suncups, and Dirt Cones}
\author{M. D. Betterton}
\address{Department of Physics, Harvard University, Cambridge, MA 02138}
\maketitle

\begin{abstract}
Penitentes and suncups are structures formed as snow melts, typically
high in the mountains. When the snow is dirty, dirt cones and other
structures can form instead. Building on previous field observations
and experiments, this work presents a theory of ablation morphologies,
and the role of surface dirt in determining the structures formed. The
glaciological literature indicates that sunlight, heating from air, and
dirt all play a role in the formation of structure on an ablating snow
surface. The present work formulates a mathematical model for the
formation of ablation morphologies as a function of measurable
parameters. The dependence of ablation morphologies on weather
conditions and initial dirt thickness are studied, focusing on the
initial growth of perturbations away from a flat surface.  We derive a
single-parameter expression for the melting rate as a function of dirt
thickness, which agrees well with a set of measurements by
Driedger. An interesting result is the prediction of a dirt-induced
travelling instability for a range of parameters.
\end{abstract}
\newpage

\begin{figure}[!hb]
\centerline{\epsfysize=2.5in\epsfbox{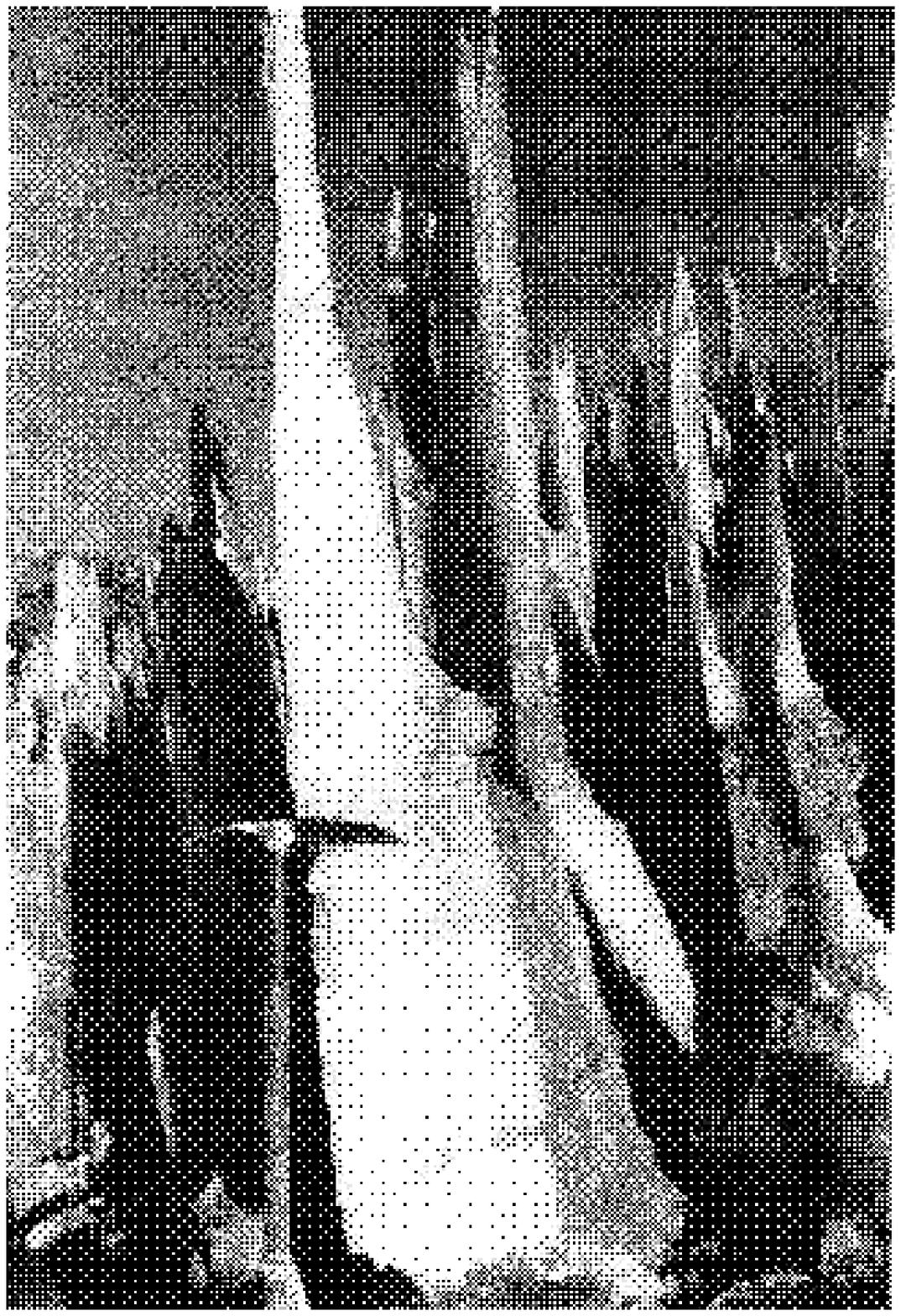}
\epsfysize=2.5in\epsfbox{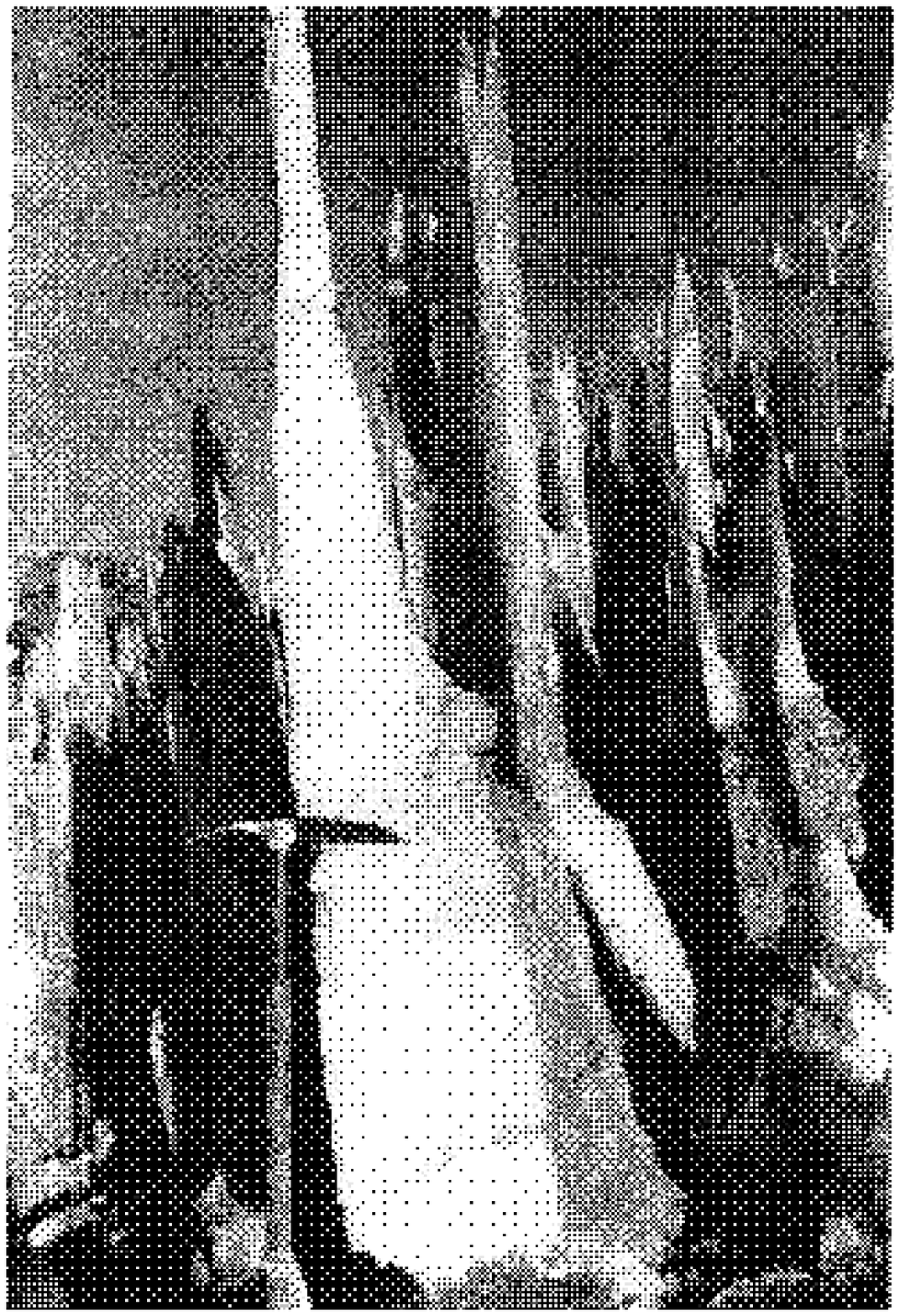}}
\caption[Penitentes]
{Photographs of penitentes, from Post and LaChapelle \cite{pos71},
p. 72. Left, penitentes on Cerro Negro, Chile. Right, field of
penitentes, north slope of Cerro Marmolejo Norte, Chile. Note the ice-axe,
approximately 80 cm high. In the picture on the left, the snow in the
hollows has completely melted, exposing the soil underneath. This is a 
frequent, though not universal, feature of penitentes \cite{rho87}.}
\label{peni1}
\end{figure}

\begin{figure}[!b]
\centerline{\epsfysize=2.25in\epsfbox{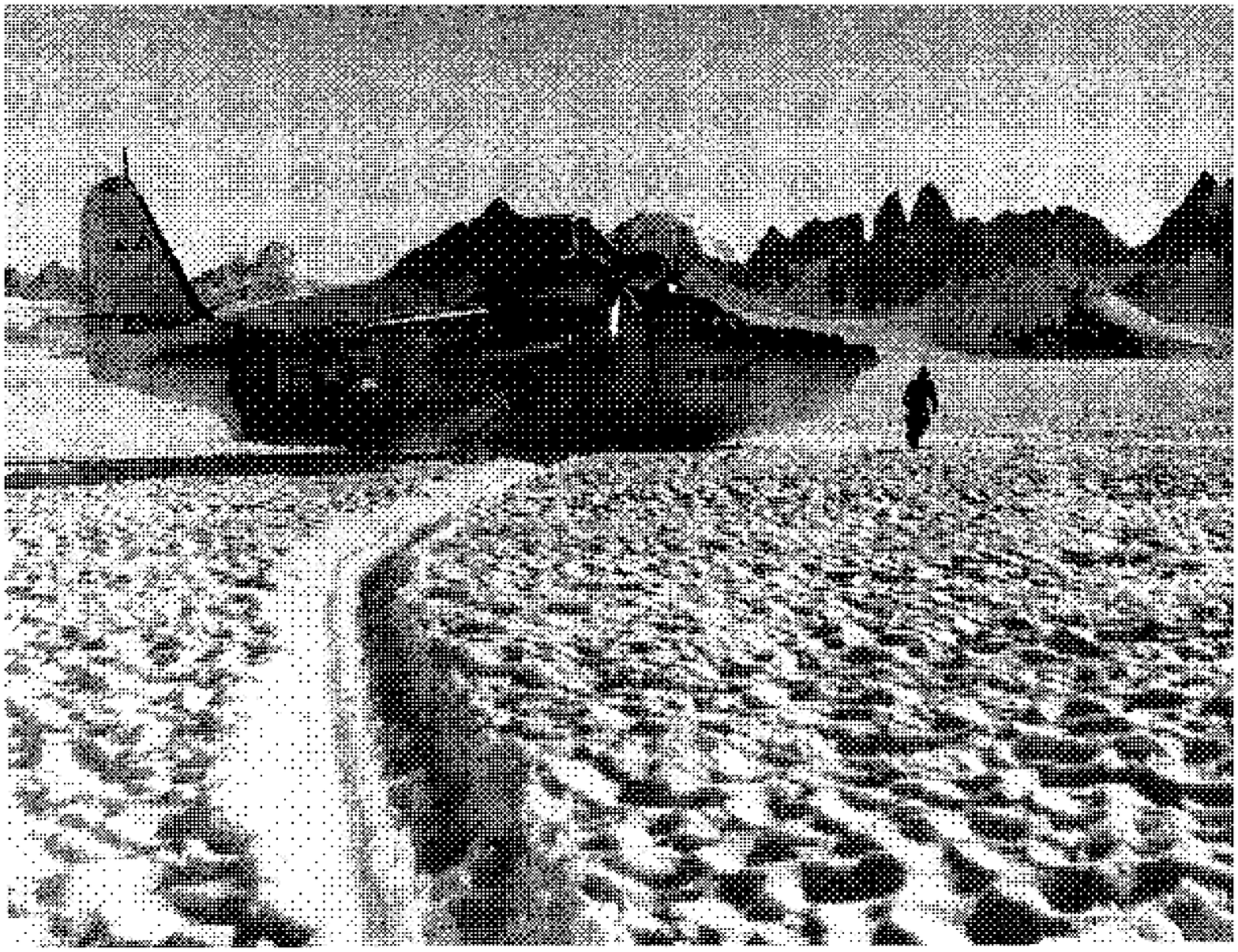}\epsfysize=2.25in\epsfbox{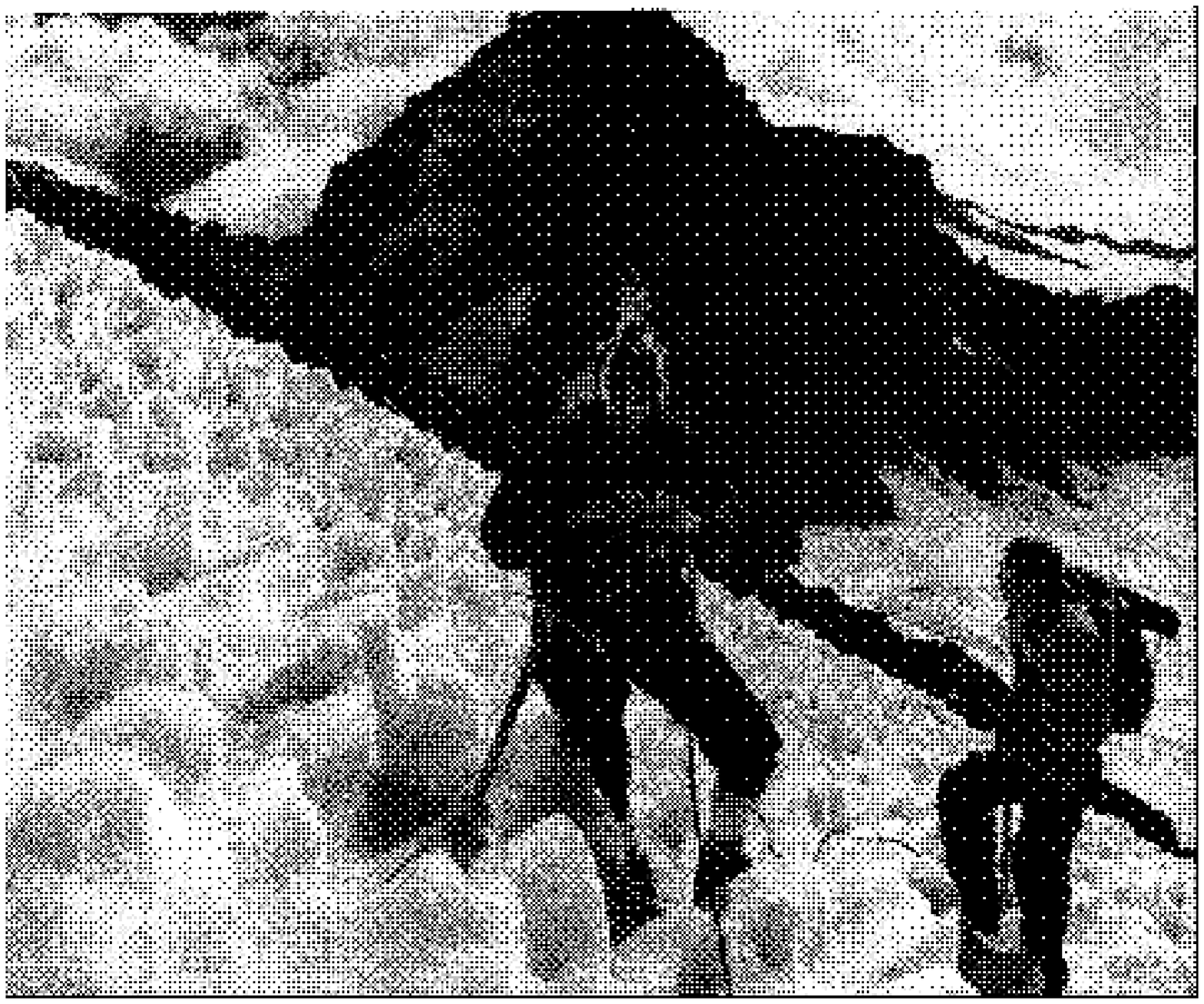}}
\caption[Suncups]
{Photographs of suncups, from Post and LaChapelle \cite{pos71}. Left,
suncups on the Taku Glacier, Coast Mountains, Southeast Alaska,
p. 70. Right, deep suncups in Disappointment Cleaver, Mount Rainier,
p. 71.}
\label{cups1}
\end{figure}

Penitentes are structures of snow or ice \cite{pos71}, which commonly
form during the summer on glaciers or snow fields at high altitudes
(in the Andes and Himalaya).  A penitente is a column of snow, wider
at the base and narrowing to a point at the tip.  The name
``penitente'' is a Spanish word meaning ``penitent one,'' and arose
because a field of penitentes resembles a procession of monks in white
robes.  Penitentes range from one to six meters high with the spacing between
columns comparable to their height (Figure \ref{peni1}).  Smaller
structures, known as suncups or ablation hollows, can be found in
lower mountains like the Rockies and the Alps (Figure
\ref{cups1}). Suncups are smaller, two cm to half a meter in size.

The first written record of penitentes comes from Charles Darwin, who
observed them during his travels in the mountains of Chile \cite{dar36}. 
\begin{quotation}
``Bold conical hills of red granite rose on each hand; in the valleys
there were several broad fields of perpetual snow. These frozen
masses, during the process of thawing, had in some parts been
converted into pinnacles or columns, which, as they were high and
close together, made it difficult for the cargo mules to pass. On one
of these columns of ice, a frozen horse was sticking as on a pedestal,
but with its hind legs straight up in the air. The animal, I suppose,
must have fallen with its head downward into a hole, when the snow was
continuous, and afterwards the surrounding parts must have been
removed by the thaw.''
\end{quotation}

An extensive literature of observations and field experiments has
documented these ablation morphologies (see \cite{rho87} for many
references).  Ablation in this context means removal of snow by
melting or sublimation. This contrasts with other processes like wind,
avalanches, and rain.  There is a consensus about the causes of
ablation morphologies, although some contradictory claims do exist in
the literature. For penitentes, bright sunlight and cold, dry weather
are apparently required \cite{rho87}, while ablation hollows can be
formed in three distinct ways, with solar illumination important in
some settings. For other locations, uniform heating from the air
appears to be the key effect. The effect of this ``sensible'' heat
transfer (so called because it is easily felt with the senses) to the
snow depends on whether the snow is clean or dirty. Since many readers
are likely to be unfamiliar with the glaciological literature, I give
a brief review here.

The observational evidence for sunlight-driven formation of penitentes
is abundant.  In early work, Matthes\cite{mat34} argued that a variety
of ablation forms, from sun cups a few inches in size to penitentes
many feet deep, are formed by the sun.  As he pointed out, the
formation of the largest penitentes requires strong and prolonged
solar radiation---the primary reason why penitentes develop only in
regions with dry summer climates.  Matthes also observed that
penitentes tilt toward the elevation of the midday sun (an observation
confirmed by others
\cite{pos71,has81,kot74,tro42,lli54}). Such tilting is strong evidence
that the sun has an important role in the development of structure,
because the direction of incident radiation provides the symmetry axis
in the problem.  In later work, Lliboutry \cite{lli54} observed that
incipient penitentes begin as east-west rows.  Perhaps most important,
if the weather is not dominated by direct sunlight---if the weather is
cloudy\cite{mat34} or very windy\cite{mat34,lli54}---penitentes are
observed to decay.  In the 1930s Troll performed an experiment trying
to create penitentes\cite{tro42}.  The exact statement (reported by
Lliboutry\cite{lli54}) is ``Troll was able to reproduce penitents in
Germany by shining an electric bulb on fresh snow during a cold, dry
night.'' This supports the sunlight mechanism, although to my
knowledge no controlled laboratory experiments have investigated
light-driven structure formation.

To understand qualitatively how sunlight can cause structure formation,
note that when light is reflected off the snow, the base of a
depression receives more reflected light than the neighboring peaks.
This drives an instability of the surface and the amplitude of a
perturbation grows; quantifying this argument will be a main goal of
this paper. The effects of reflections are considered imporant by
several observers \cite{mat34,lli54,ams58}. This may not be
the only required effect. At the high altitudes where penitentes
commonly form, the air is so cold and dry that sublimation occurs
instead of melting \cite{ben98}, consistent with the observations that 
the snow in penitentes is quite dry \cite{mat34,lli54}. Lliboutry
\cite{lli54} claims  that the snow in the hollows between penitentes is soft
and wet, and that temperature variations of 5-10$^{\circ}$ C exist
between the peaks and the troughs. This was interpreted to indicate
snow sublimating from the peaks and melting near the troughs---an
effect which accelerates the growth of structure, since 7 times more
heat is required to sublimate a volume of snow than to melt
it. Lliboutry believes this effect to be crucial for the development
of the largest structures, and claims that penitentes only appear at
altitudes high enough that sublimation becomes important. But other
researches report results in disagreement with this
\cite{mat34,kot74}.

A different set of observations and experiments have led to a very
different claim: that solar illumination destroys ablation
morphologies, while windy weather promotes their growth.
Leighly\cite{lei48} argued that heat from air (delivered by wind)
leads to the formation of ablation polygons (cf. Figure 3).  Others
state \cite{ric54,ric57,jah68} that structures do not grow in the
presence of direct sunlight.  Ashwell and Hannell claim\cite{ash66}
that when the incident solar power is larger than the incident power
from heating by wind the hollows decay. Detailed observations, along
with wind-tunnel experiments, have been made by Takahashi and
collaborators\cite{tak73,tak78}; they conclude that structures grow
most rapidly when the air temperature and wind speed are
highest\cite{langbarr}.  When the weather is warm and cloudy, wind
mixes the air so heat is delivered at a steady rate to the surface;
the higher the temperature and wind speed, the faster the
heating. 

\begin{figure}[!b]
\centerline{\epsfxsize=3in\epsfbox{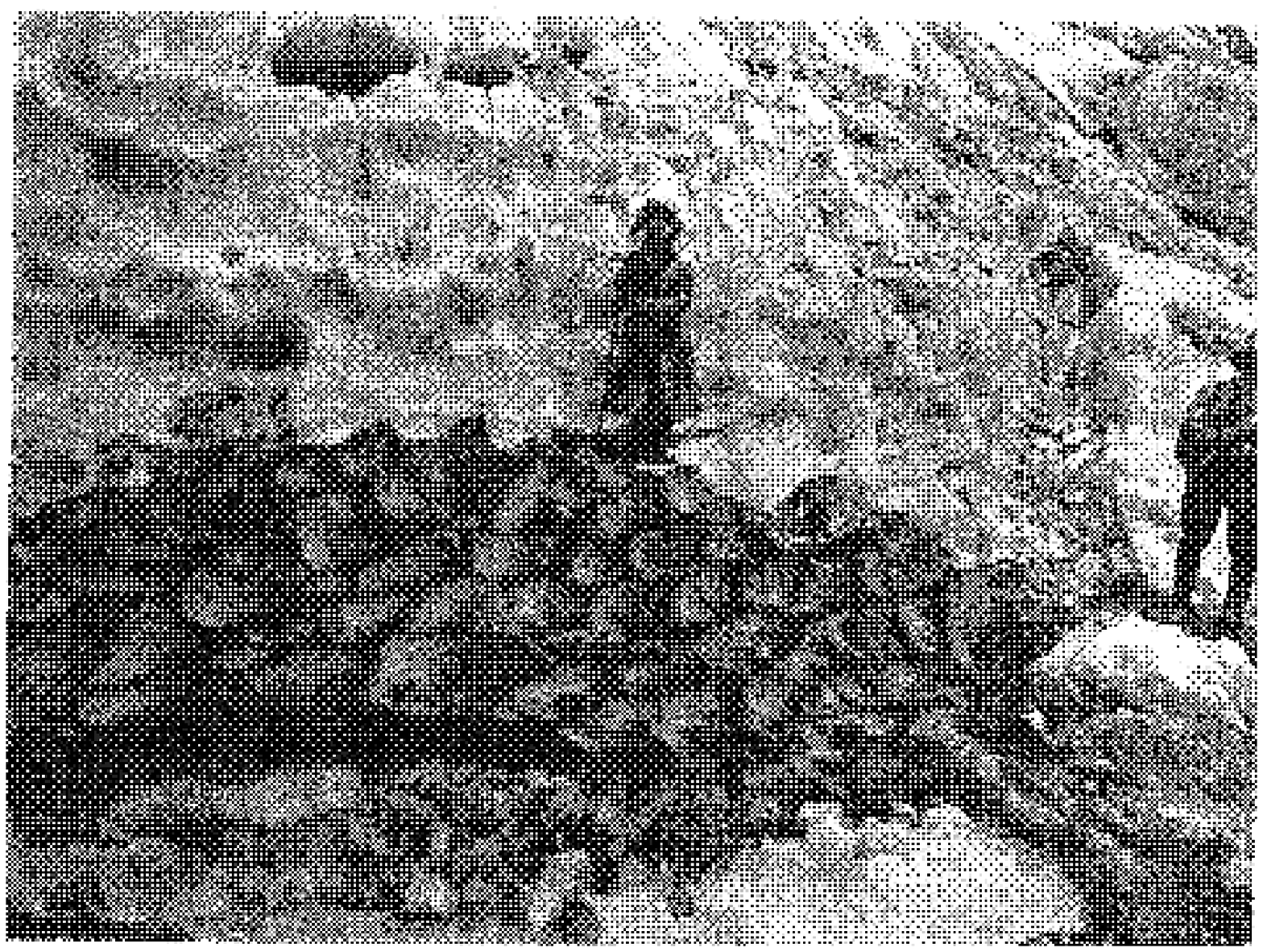}\epsfxsize=3in\epsfbox{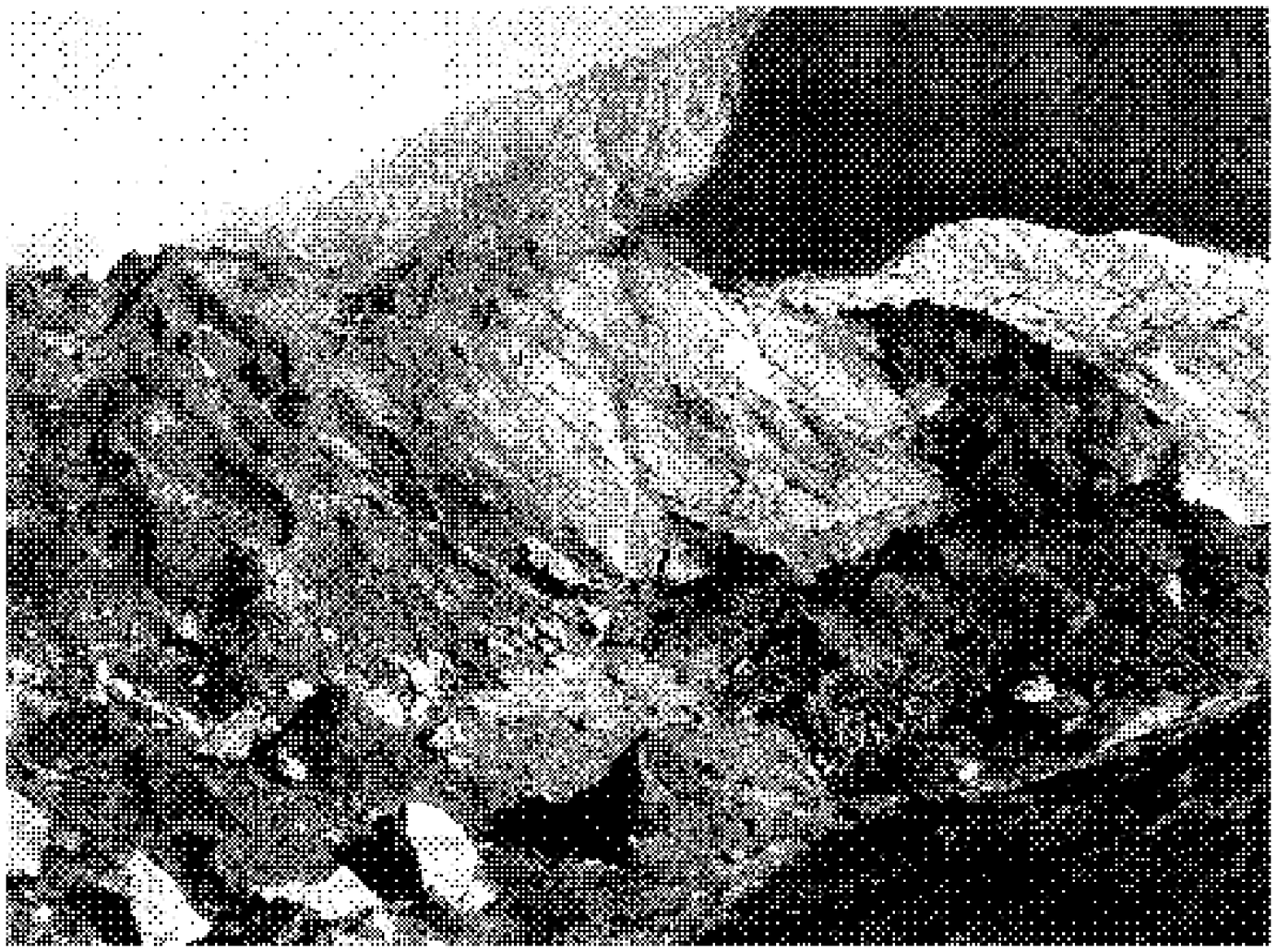}}
\caption[Dirt-Driven Structures]
{Photographs of dirt-driven structure, from Workman and Workman
\cite{wor09}. Left, ablation hollows with dirt collected on the
ridges. The structures are reportedly 12 to 18 inches high,
p. 196. Right, dirt cones, approximately 20 to 40 inches high, p. 190.}
\label{dirtpic}
\end{figure}

Rhodes, Armstrong, and Warren \cite{rho87} suggested a resolution to
this apparent contradiction, which I now summarize.  In their view, dirt on the snow surface is the hidden
variable distinguishing the two cases.  Sunlight drives formation of
penitentes in clean snow because reflection into hollows makes depressions
in the snow surface grow.  Any source of ablation which transfers heat
uniformly to the snow surface therefore disrupts formation of
structure.  However, sunlight acts differently on a {\sl dirty} snow
surface.  Dirt decreases the amount of reflected light, preventing the
concentration of sunlight in the hollows.  This agrees with the Rhodes
{\it et. al.} observations of suncups on Mount Olympus.  The
researchers noticed that when the snow surface was covered by a layer
of ash from the eruption of Mount Saint Helens, suncups did not form.
They scraped away the ash from one patch of snow and observed the
formation of sun cups on this clean snow surface.

How does dirt affect snow ablation?  If the dirt thickness covering the
snow is sufficiently thick, the dirt forms an insulating layer which
slows down the ablation rate of the snow \cite{wil53}.  Thus dirt can
have different effects, depending on thickness.  A thin layer of dirt
causes faster ablation because reflection is inhibited.  However,
sufficiently thick dirt slows ablation.  A large amount of work has
looked at how debris-covered ice or snow melts
\cite{ben98,ash66,dre72}; one typically finds a peak in the
ablation rate for dirt thickness around 0.5--5 cm. One nice experiment
was done by Driedger\cite{dri81}, who measured ablation rate as a
function of ash thickness on the South Cascade Glacier (primarily due
to melting).  The typical grain sizes of the ash were 0.25 to 1.0 mm
diameter, and the maximum ablation rate occurred for a dirt thickness
of 3 mm.  The data from her measurements are shown in Figure
\ref{dirtmelt}. Comparison to these data provides a test of my model,
as discussed below.

\begin{figure}[!t]
\centerline{\epsfysize=1.5in\epsfbox{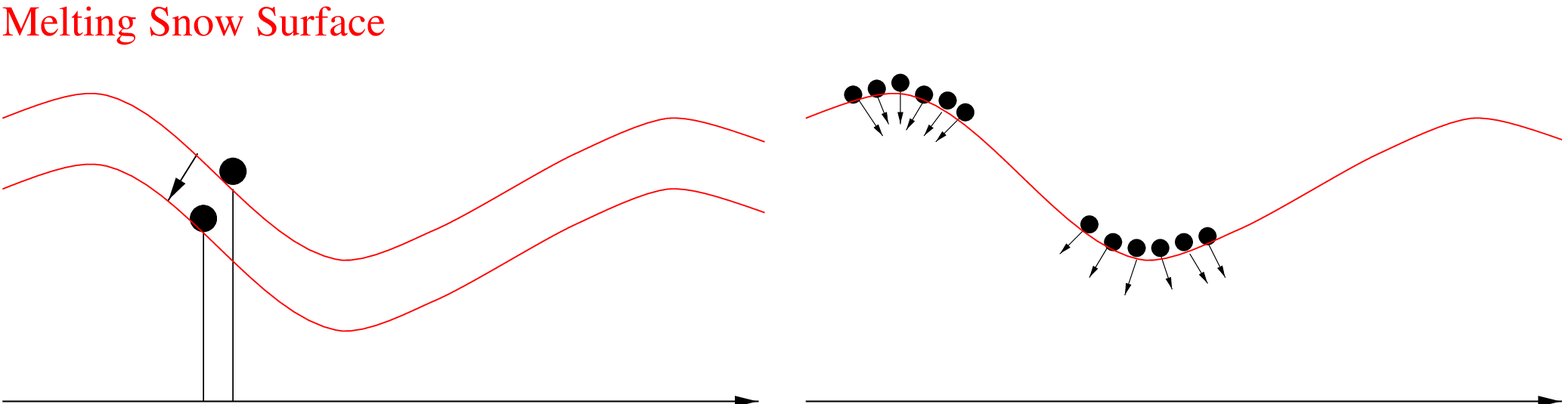}}
\caption[Schematic of Dirt Motion on Snow Surface]
{Motion of dirt on a snow surface. A particle adhered to the surface
of the snow moves normal to the surface (left). When particles follow
such ``normal trajectories,'' peaks are stable equilibria and valleys
unstable equilibria (right).}
\label{dirt}
\end{figure}

As pointed out by Ball\cite{bal54}, small particles of dirt adhere to
the snow surface. This is true only for sufficiently small dirt
particles: the adhesive force on the particle must be large compared
to the gravitational force\cite{ash66}.  When adhesion to the snow
dominates, the pieces of dirt move perpendicular to the snow surface
(rather than falling straight down) as the snow ablates.  Sticky dirt
therefore tends to become concentrated on the most elevated regions of
the surface. (See Figure \ref {dirt}.) The concentration of dirt on melting
snow can be observed in old snow piles in cities, and is illustrated in Figure
\ref{dirtpic}. This movement of dirt normal to the melting snow
surface is quantitatively well-documented in the
literature (summarized in \cite{rho87}). For the arguments here to be
correct qualitatively, the dirt need not move completely normal to the 
surface---a component of motion normal to the surface is adequate.

This mechanism explains dirt-driven structure formation: as the snow
ablates, dirt becomes concentrated on the more elevated parts of the
surface.  The thicker dirt forms an insulating layer on the ridges, so
they ablate more slowly.  The hollows thus grow deeper. This
concentration of dirt by ablation can, in extreme cases, lead to the
formation of so-called dirt cones: cones of snow or ice covered by a
thick layer of dirt\cite{ben98,wil53,dre72,swi50}. (See Figure
\ref{dirtpic}.) These structures can become quite large: Swithinbank
\cite{swi50} reports a dirt cone in the Himalaya estimated to be 85 m
high! Drewry \cite{dre72} has done detailed experiments on dirt
cones. He concludes that the cones ultimately reach a steady state,
where the motion of dirt toward the center of the cones is balanced by
the debris sliding down the cone when the slope angle is too large.

The proposal of Rhodes, Armstrong, and Warren \cite{rho87} that
uniform heating causes structure only for dirty snow does not
completely resolve the disagreement about structure formation. Some observers
who advocate uniform-heating driven formation of ablation hollows
insist that dirt on the snow surface is not required
\cite{lei48,ric57,tak73,tak78}. Indeed, some photographs show ablation 
hollows in clean snow inside a tunnel or on other inverted surfaces,
suggesting that neither dirt nor solar illumination are necessary. How
can this be explained? Some have suggested that a regular pattern of
convection cells leads to the observed polygonal
pattern\cite{lei48,ric54}, but a simple estimate shows this cannot
give the correct size structures \cite{nod00}. Another suggestion is
that the structures are formed by turbulent eddies\cite{ric57,tak73},
although Takahashi \cite{tak78} later claimed that the diameters of
the hollows are independent of the eddy size. Takahashi \cite{tak78}
proposed that the separation of the air boundaray layer as it flows
over a cusp could produce lower temperatures at the cusp, and
therefore lead to structure formation. I am not familiar with any
further theoretical or experimental consideration of the Takahashi
proposal; this mechanism for structure formation will not be
considered in this paper.

Despite the extensive observations of ablation morphologies, there is
a lack of mathematical models of their growth \cite{theorynote}.  The
goal of this paper is to quantify the primary mechanisms discussed
above, and characterize the initial stages of the instability of a
flat surface. We examine how reflection of sunlight and dirt affect
structure formation. The nonlinear evolution more appropriate to
penitentes will be examined in a future paper.

In this paper we consider sunlight---direct and reflected---the primary
source of heat which leads to snow ablation. It is well-documented
that radiation is the dominant heat source for ablating snow
\cite{wis80,van99}, especially at high altitudes and low latitudes
\cite{kot74}. The importance of considering {\it reflected} light as well as
direct illumination is supported by the observational
evidence. The fraction of light reflected from old snow is about 0.5
\cite{ben98,wis80}. Therefore the amount of heat absorbed locally (and
correspondingly the ablation rate) can vary by up to a factor of two
for different parts of the surface---such a large variation can have
important consequences for structure formation. Kotlyakov and Lebedeva
\cite{kot74} made measurements of the albedo on a glacier with
small penitentes. In a measurement averaged over surface features, ten
percent more light was absorbed when the sun was high overhead,
presumably indicating the absorption of reflected light in the
structures.

In the presence of dirt, sensible heating from the air may be
important, in addition to sunlight. In this paper I focus primarily on
the sunlight-dominated case, and comment on similarities and
differences with sensible heat. Modifying the model to include
sensible heating is straightforward.

By forming a quantitative model, we can test whether the effects
considered can explain the appearance of structure, and describe the
morphologies produced.  The primary goal is to formulate the simplest
model which contains the essential physics. Ideally the theory would
contain no free parameters, that is, all parameters in the model can
be calculated or measured in experiments. We also discuss which
effects are left out of the simple model, and estimate how serious are
the consequences for such omissions.

The first part of this paper addresses clean snow only. In section
\ref{modelclean} we formulate a minimal model, and carry out the
analysis of the model for small perturbations. The linear stability
analysis lets us estimate the wavelength of the fastest-growing
disturbance, and determines the initial size structures that form.
This analysis allows us to make testable predictions of the conditions
necessary to observe structure.

We then discuss the effect of dirt and re-formulate the model to
include dirt in section \ref{modeldirt}. We compare our model to the
field experiment of Driedger and find good agreement. Thus reassured
that the theory contains the imporant physical effects, we show how
dirt alters the growth of small perturbations.  We show that a thin
dirt layer suppresses the reflection-driven instability and induces
travelling dispersive waves on the surface. In the limit of thick
dirt, we demonstrate the insulation-driven instability expected from
the discussion above.

\section{Light Reflection on Clean Snow}
\label{modelclean}

The model for penitente growth we derive here contains simplifying
assumptions; we hope to capture the essential features while
neglecting some effects We will discuss the assumptions and their
limits of validity as the model is developed. Some of the most
important simplifications include considering the latent heat to be
constant and including only first-order, isotropic reflections.  We
focus on a one-dimensional model of penitentes, assuming invariance in
the transverse direction, although it is straightforward to generalize
these equations to two dimensions or to multiple reflections.

We consider the height of the snow surface $h (x, t) $, and seek an
equation for the time evolution of $h$.

\subsection{Snow Ablation}

Heat incident on the surface leads to ablation---the height $h$
decreases as the snow melts or sublimates.  We assume that ablated snow
vanishes into the air or drains, and therefore that the flow of water along
the surface and re-freezing are not important (and similarly that other
changes in the nature of the snow are unimportant). This model can
apply to either melting or sublimation. We use the term ``ablation''
to refer to removal of snow in either way.

Suppose a point on the surface absorbs a power per unit horizontal
area $P(x)$.  The latent heat required to ablate a unit volume of snow
is $L$. Combining this with an effective diffusive smoothing term (see
below) gives the evolution equation for the surface:
\begin{equation}
\frac{\partial h}{\partial t} = -\frac{P (x)}{ L } + D
\frac{\partial^2 h}{\partial x^2}.
\label{maineq}
\end{equation}

For clean snow, we assume that $L $ is a constant (independent of $x
$). This is true when the surface temperature and humidity are
approximately constant.  As discussed in the introduction, fully
developed penitentes may have melting in the hollows and sublimation
in the tips---a situation which requires $L $ to vary along the
surface. Indeed, the variation in $L $ might be the essential effect
for large structures. For small angle structures, that is, amplitude
small relative to wavelength, $L = $ constant should be a good
approximation. Later, we will include spatial variation in the
effective $L $ due to dirt on the snow surface---see section
\ref{modeldirt}.

The second term in equation \ref{maineq} for the surface height is a
simple form of the small-scale cutoff: a diffusive term with diffusion
constant $D $.  As we will see below, in the absence of any smoothing
term, the model can produce arbitrarily small structures.  This is
clearly not realistic, because the physics at small scales will cut
off the instability. For the qualitative results here, the exact
mechanism of the small-scale cutoff is not essential; the main point
is that there is some minimum size structure which can form. A natural
small-scale cutoff is the extinction length of sunlight, which defines
the thickness of the snow layer in which the light scattering takes
place\cite{wis80}.  Points on the snow surface within one extinction
length are not optically independent, and therefore such nearby points
ablate at the same rate.  The extinction length depends on the density
and grain structure of the snow. The typical extinction
length\cite{wis80,van99} for old snow (grain radius 1 mm) is of order
1 cm\cite{lightnote}. We will choose the diffusion coefficient so that
the characteristic cutoff length is of order the optical extinction
length. Again, remember that this term in the height equation is a
simplified representation of the small-scale physics, and any
conclusions which depend sensitively on the form of this term should
be considered suspect. Note that diffusion of heat
through the snow might seem another natural form of the small-scale
cutoff; however the gradients of temperature in the snow are not large 
enough for thermal diffusion to stabilize short wavelengths\cite{heatdiff}.

We note here that recent work by Nodwell and Tiedje \cite{nod00}
considers the scattering of light in the snowpack in quantitative
detail. They find a range of length scales where suncups can form
(both a minimum and maximum wavelength), a result which our simplified
model cannot produce.

\subsection{Light Reflection}

In this section we describe the reflection of sunlight from the snow surface.
We assume that the sunlight shines directly down (in the $-z$
direction) and has a uniform power per unit length $I$. The parameter
characterizing reflections is the albedo $\alpha$, which denotes the
fraction of light {\sl reflected}. Thus the absorbed power per unit length is
$(1-\alpha) I$. For old snow---called firn---a typical value is $\alpha = 0.5$ 
\cite{ben98}.

The reflecting properties of snow are different from those of a
mirror. Snow looks white because it scatters light in many directions,
as we would expect for a rough surface. Here we treat the light using
ray optics, and assume the surface reflects isotropically, thus the
power is distributed uniformly into $\pi$ of solid angle outside the
surface. We approximate that the reflection occurs at the surface of
the snow. (As mentioned above, the reflection takes place in a layer
of order 1 cm thick. We ignore this in formulating the reflections,
and include its effects schematically through the diffusive term.)

\begin{figure}[!b]
\centerline{\epsfysize=3in\epsfbox{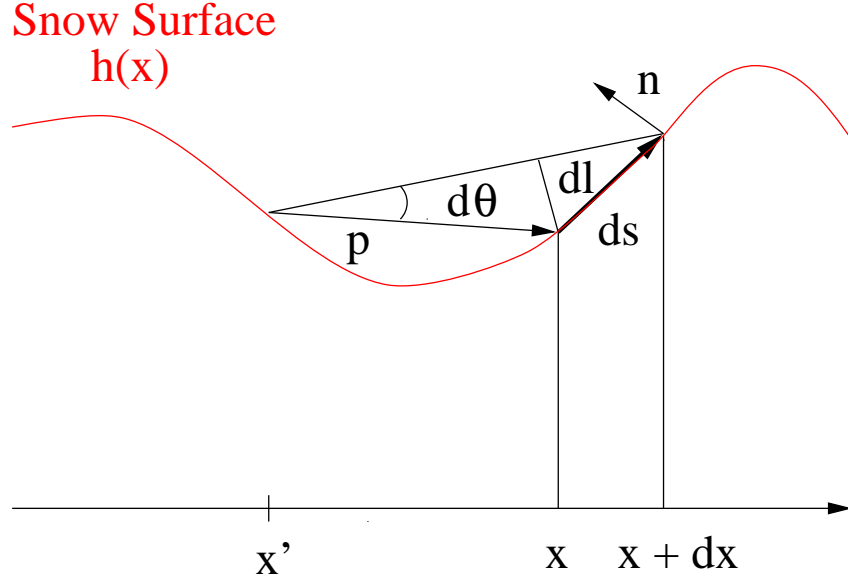}}
\caption[Schematic of Ablating Snow Surface]
{Schematic of the ablating snow surface. Scattering from the point $x_1$ 
to the interval between $x$ and $x+dx$ depends on the angle $d
\theta$. The vector ${\bf p}$ points from $x_1$ to $x$ and the increment
$d l$ is normal to ${\bf p}$ such that $d \theta= dl/p$. The vector
${\bf n}$ is normal the surface at $x$ and $d {\bf s}$ is the
increment along the surface between $x$ and $x+dx$.}
\label{meltsketch}
\end{figure}

Using these properties, the total amount of light scattered from an
interval around point $x_1 $ to the interval between $x $ and $x+dx $
is
\begin{equation}
 \frac{\alpha I}{ \pi} d \theta \ dx_1 ,
\end{equation}
where $d \theta$ is the angle subtended by the surface between $x$ and
$x+dx$ (see Figure \ref{meltsketch}).  

We can find $d \theta$ in terms
of the shape of the surface.
\[
d   \theta =\frac{dl}{p} =\frac{|{\bf p} \times d{\bf s}|} {p} =\frac{
\Delta h- \Delta x h' (x)}{  \Delta h^{ 2} +\Delta x ^{ 2}}.
\]
where we have used $h' = \partial h/\partial x$ and
\begin{eqnarray}
\Delta
x & = & x_1 -x\\
\Delta h & = & h (x_1) -h (x)
\end{eqnarray}
and $d{\bf s} $ is the vector tangent to the surface 
\begin{equation}
d{\bf s} = dx (1, h') 
\end{equation}
We define the vector ${\bf p}$, which points from the point $x_1 $ to
the point $x$.  From Figure \ref{meltsketch}, we can see that
\begin{equation}
p=  \sqrt{ \Delta x ^{ 2} + \Delta h^{ 2} } 
\end{equation}

To find the total power reflected to point $x$, we must add up the intensity 
scattered from all points $x_1$:
\begin{equation}
P_r(x) =  \frac{  \alpha I}{ \pi} \int \frac{dx_1 (\Delta h - h'(x)
\Delta x) } { \Delta x^2 + \Delta h^2 } 
\end{equation}
The integrand in this equation is the propagator for light intensity,
it describes how the intensity is carried from one point to another on
the surface; $P_{ r} (x) $ is the intensity due to a {\sl single}
reflection.  To include multiple reflections, we can write the power
as an integral equation for $P $.
\begin{equation}
P(x)= (1-\alpha) I + \frac{ \alpha }{ \pi} \int \frac{dx_1 P(x_1) (\Delta h - h'(x) \Delta x)} { \Delta x^2 + \Delta h^2 }
\end{equation}
This can be written as a power series in $\alpha$. We will only
consider single reflections here, which does not 
introduce a large error when $\alpha $ is small. For old snow, a
typical value of $\alpha \approx 0.5 $. Including the higher-order
correction from multiple reflections 
may be important in determining the precise details of the largest
shapes. 

This formula for reflected intensity is not complete, because it
neglects the {\sl line-of-sight} constraint. Light cannot scatter from
$x_1$ to $x$ if the path of the light ray is blocked by another part of
the surface. This requirement is a nonlinear constraint which is
difficult to handle analytically but is straighforward to implement in 
numerics. We typically indicate the constraint schematically, by writing
``line of sight'' under the integral:
\[
P_r(x) = \frac{  \alpha I}{ \pi} \int_{  {\tiny \begin{array}{c} 
	\mbox{ line of} \\ \mbox{ sight} \end{array} } }
\frac{dx_1 (\Delta h - h'(x) \Delta x)} 
	{ \Delta x^2 + \Delta h^2 }
\]
We can also write a necessary (but not sufficient) criterion for the
line of sight constraint, when applied to local analysis within one
``basin.''  The two points $x$ and $x_1$ are within line of sight of
each other when the dot product of the vector normal to the surface
and the vector ${\bf p} $ is less than 0: $- {\bf n} \cdot {\bf p}
=\Delta h- \Delta x h' (x) > 0 $. (See Figure \ref{meltsketch}.) Note,
however, that this simple criterion will miss intermediate bumps in
the surface.  In other words, the constraint may be satisfied but no
reflection occurs between $x_1 $ and $x $ because the line of sight is
blocked by an intervening peak.

\subsection{Model}

The equations combining reflection and ablation are
\begin{equation}
\frac{\partial h}{\partial t}  =    - \frac{\alpha I}{L}{\cal I} (x) + D h''
\label{evolve}
\end{equation} 
where we have defined the integral
\begin{equation}
{\cal I}(x) = \frac{1}{\pi} \int_{ {\tiny \begin{array}c \mbox{ line
of} \\ \mbox{ sight} \end{array} } } \frac{dx_1\ ( \Delta h - h'(x)
\Delta x) } { \Delta x ^{ 2} + \Delta h^{ 2}}.
\end{equation}
The intensity of the sun determines a characteristic ablation rate $I L^{-1}$,
where $L$ is the latent heat per unit volume. Combining
this velocity with the diffusion coefficient $D$ gives a length
\begin{equation} 
\bar \ell= \frac{D L}{I}
\end{equation} 
and time 
\begin{equation} 
\bar t=\frac{D L^2}{I^2}.
\end{equation} 
The solar constant
gives the intensity of solar radiation at the top of the atmosphere
\cite{van99} $I=1.4 \times 10^{6} \ \mbox{erg cm}^{-2}\
\mbox{sec}^{-1}$; we therefore choose $I=10^{6}\ \mbox{erg cm}^{-2}\
\mbox{sec}^{-1}$ as the typical value of $I$ under bright sunny
conditions. The latent heat depends on density. Freshly fallen snow
has a density of between 0.05 and 0.2 $\mbox{g cm}^{-3}$, while older
snow that has survived one melt season has a density range of 0.4 to
0.8 $\mbox{g cm}^{-3}$ \cite{ben98,lli54}. Here we pick an
intermediate density of 0.3 $\mbox{g cm}^{-3}$ for our estimates. This
gives a latent heat per unit volume for melting $ L = 10^9 \ \mbox{erg
cm}^{-3}$ and a melting rate $I/L = 10^{-3}\ \mbox{cm
sec}^{-1}$\cite{meltnote}. We pick $D=2.5 \times 10^{-5}\ \mbox{cm}^{2}\
\mbox{sec}^{-1}$, where this choice is made so that the most unstable
wavelength is 2 cm (see below). In this case the length scale $\bar
\ell = 0.25$ mm. and the time scale $\bar t = 25$ seconds.

For sublimating snow, the latent heat is seven times larger. this
gives the slower melting rate $I/L=1.4 \times 10^{-3}\ \mbox{cm
sec}^{-1}$, larger length scale $\bar \ell = 1.75$ mm and time scale $\bar t
= 1225$ seconds.

We will now perform a perturbation analysis of
Equations \ref{evolve} to see how the size structures formed
compares to the scale $\bar \ell$. We have set up the problem so that
structures will initially form on a scale roughly comparable to
$\bar \ell$, and expect the perturbation analysis to give this result.

\subsection{Quasi-Linear Regime}
\label{linclean}

Here we show how an approximate linear analysis of the equations can
be performed.  This allows us to derive the dispersion relation, which
characterizes when the system is stable or unstable.  There is a
fastest growing mode determined by the competition between reflection
and diffusion. The length scale of this mode is related to the
basic scale $\bar \ell$ from dimensional analysis  above; we determine the prefactor
here. The results are significant because they describe how
the physical parameters affect the instability. We will argue that
reflection favors structures on scales as small as possible. On the
other hand, the small-scale cutoff limits the smallest structures
possible. Therefore we expect the fastest-growing mode to be of order
the cutoff size.

The reflection integral is scale invariant: upon
rescaling $x$ and $h$ by the same amount the integral ${\cal I} (x) $ is unchanged. Thus in
the absence of diffusion, there is no characteristic scale in the
problem. Therefore a shape with aspect ratio 1---a shape with
variations in $h $ comparable to variations in $x $---should have a
growth rate of order 1 (in the absence of boundary effects). The integral
contributes a shape factor independent of the amplitude of the shape
$\delta $. Therefore the rate of change of amplitude $\dot{\delta } $
is constant.

To examine shapes with aspect ratio far from one,  we start with an
aspect-ratio 1 shape, then transform $x \rightarrow \lambda x$ and $h
\rightarrow \delta h$ . When  $\delta \ll \lambda$, we find that the
integral scales with the basic angle $\delta/\lambda$: ${\cal I}
\rightarrow \delta/\lambda {\cal I}$. Thus for small perturbations, we
expect a growth rate proportional to the amplitude ($\dot{\delta} \sim
\delta$). 

For sufficiently small $\delta/\lambda$, we treat the contribution
from the reflection integral as a numerical factor of order 1. Note
that a sinusoidal perturbation is not an eigenshape for small
amplitudes; we do not know what the actual eigenshapes are. The dominant
contribution is the scaling with $\delta $, and we neglect the other
(slower) dependence on position, amplitude, etc. Thus the {\sl
quasilinear} equation for a small-amplitude variation in the surface
$h = \delta \sin qx \ e^{\omega t}$ is approximately
\begin{equation}
{\cal I}(x)  \approx  \frac {q} {\pi }  \delta \sin qx \ e^{\omega t}
\end{equation}
which gives a dispersion relation
\begin{equation}
\omega  =  \frac {\alpha  I} {\pi L} q - D q^2
\label{quasilin}
\end{equation}
This argument selects a  fastest-growing mode with wavenumber
\begin{eqnarray}
q_* &=& \frac{\alpha I }{2 \pi L D}	\\
\omega_* &=& \frac{(\alpha I )^2}{4\pi^{ 2} L^2 D} = q_*^2 D
\end{eqnarray}
These equations are the dimensional analysis result, with an estimate
of the prefactor from the scaling argument. Plugging in values of
typical parameters given above, we find the most unstable wavelength
for melting $\lambda_*=2 \pi/q_*$ of 2 cm, and characteristic time
4000 sec. In the case of sublimation the wavelength is 14 cm and the
time $2 \times 10^{5}$ sec. The choice of the diffusion coefficient is
now clear: we chose $D$ to give a most unstable wavelength of 2 cm. We
have put in diffusion as a simplified representation of the
small-scale physics, and chosen its value so that the numbers make
sense. It is important to remember that because of this choice of $D$,
the numbers calculated here cannot be considered a prediction of the
initial size structures that form. The calculation of real interest is
how this instability is changed by dirt, as discussed in the following
section.

Although it agrees well with simulations of initial growh of
perturbations which compute the reflected intensity at each point , we
must remember that this analysis is only quasi-linear because we do
not know the eigenfunctions of the reflection integral, and
superposition does not hold: because the integral is nonlocal, a
surface variation with two modes of different wavelength cannot be
described by the addition of two modes with different $q$.

\section{Effects of Dirt}
\label{modeldirt}

A layer of dirt on the surface of the snow changes its properties.  We
model both the optical and insulating effects of dirt, and fit the
theory to melting data measured by Driedger\cite{dri81} . These data
allow  measurement of a crucial parameter in the model, and the good
agreement between theory and experiment show that we have captured the
important effects of dirt. The essential features are that thin dirt
speeds ablation, because it increases absorption, while thick dirt
insulates the snow, slowing ablation. This basic behavior leads to the
two different regimes of instability \cite{rho87}.

Dirt looks black because it absorbs light. The presence of dirt
effectively decreases the surface albedo and therefore increases the
fraction of absorbed light.  We assume light has a probability of
being absorbed that is constant per unit thickness of dirt. The
fraction of light not absorbed by the dirt is $e^{-s/s_e}$\cite{extnote}, where
$s$ is the dirt thickness and $s_e$ the extinction length in the
dirt---typically of order the
characteristic dirt particle size. Therefore dirt modifies the albedo
according to
\begin{equation}
\alpha_d = \alpha e^{-s/s_e} .
\end{equation}
Note that absorption by the dirt layer is not isotropic---more light
will be absorbed near grazing incidence, decreasing the reflection
even more. The qualitative effect of dirt remains the same however,
and thus we neglect this anisotropy. Increased absorption through a
lower effective albedo hastens snow ablation.

But the dirt also slows ablation. In the presence of an {\sl
insulating} dirt layer, the temperature at the surface of the snow is
decreased below the ambient temperature, and more heat is required to
ablate a given amount of snow.  Suppose an amount of heat $L$ is
necessary to ablate a unit area of clean snow.  How much additional
heat is required in the presence of a dirt layer?  At steady state the
temperature satisfies
\begin{equation}
\nabla^2 T = 0 .
\end{equation}
When the radius of curvature of the surface is large compared to the
dirt thickness (the important limit for growth of perturbations) we
can treat the snow surface as planar, leading to variations in $T$ in
the $z$ direction only.  The boundary conditions are: At the dirt-air
interface ($z=0$), the temperature must be equal to the ambient
temperature.  The temperature gradient at the surface due to heat flux
into the dirt from the air is $T' (z = 0) =P/\kappa$, where $P$ is the
incident power flux and $\kappa$ the thermal conductivity of the dirt.
Thus we find that the temperature at the snow surface is less than
$T(z=0)$ by an amount $\Delta T = P s/\kappa $.
An extra amount of heat $\Delta Q =C \Delta T $ is needed to raise the
snow temperature up to its value in the absence of dirt, where $C $ is
the heat capacity of the snow.  Thus the effective latent heat for
a dirt thickness $s $ is
\begin{equation}
 L_d =   L  +\frac{ C P s}{\kappa} 
\end{equation}
Both $ L $ and $C $ depend on the ambient temperature $T $. However,
the dependence is sufficiently weak that we can neglect it. 

\begin{figure}[!t]
\centerline{\epsfysize=3in\epsfbox{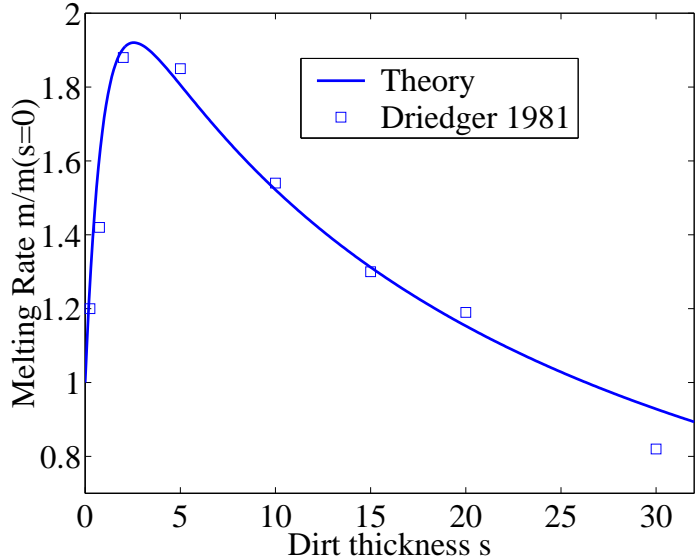}}
\caption[Ablation Rate vs. Dirt Thickness]
{ A plot of the relative ablation rate $m/m(s=0)$ versus dirt
thickness. The points are the data measured by Driedger
\cite{dri81}. The solid curve is a one-parameter fit to Equation
\ref{meltrate}, yielding the fitted $\gamma=0.047$. Note the fastest
ablation occurs for dimensionless $s \approx 3$.  We picked $s_e=1$ mm
from Driedger's measurement of the dirt particle size, and albedo
$\alpha =0.5$ from other measurements \cite{wis80}. We can also
estimate the parameter $\gamma$ (see text). The estimate gives
$\gamma$ within a factor of 2 of the value obtained from this fit.}
\label{dirtmelt}
\end{figure}

Combining these two effects we find that the snow ablation velocity
for a flat surface covered with dirt is
\begin{equation}
m(s)=\frac{I}{ L } g (s) \end{equation}
where $g $ is a dimensionless function of the dirt thickness. In this model,
\begin{equation}
g (s) =\frac{1-\alpha e^{-s/s_e }}{1 +\gamma s
(1-\alpha e^{-s/s_e})} \label{meltrate} 
\end{equation}
where we have defined the dimensionless measure of the insulating
value of dirt:
\begin{equation} 
\gamma= \frac{s_e C I}{ L \kappa}.
\end{equation} 
The non-monotonic behavior of this curve---positive slope for small $s $
and negative for large $s $---is the important qualitative result.
Note that in the absence of dirt the ablation rate is as expected:
\begin{equation}
m(s=0) =\frac{I}{ L } (1 -\alpha ) 
\end{equation}
A fit to the data of Driedger \cite{dri81} is shown in Figure
\ref{dirtmelt}. Driedger measured melting rates of a flat surface
for different dirt thickness. The plot shows $m/m(s=0)$ versus
dimensionless dirt thickness. We picked $s_e=1$ mm from Driedger's
measurement of the dirt particle size. Fitting the data to Equation
\ref{meltrate} allows us to determine  the
dimensionless insulation coefficient $\gamma$. For $\alpha \approx 0.5
$ (from other measurements \cite{wis80}) the fit gives $\gamma=0.047$.
We can also estimate the parameter $\gamma$ using other data (see
below). The estimate gives $\gamma$ close to the value
obtained from this fit.

This model and the experiment of Driedger are in the regime where
solar radiation is the dominant heat source. The discussion of Rhodes
{\it et.\ al.}\ \cite{rho87} points out that the ablation curve changes
when sensible heating is important. In fact, if radiation is
negligible the curve will monotonically decrease as the dirt thickness 
increases, because light absorption effects disappear in this
limit. It is straightforward to adjust the model to include other
sources of heating. Measurements of the type Driedger performed,
compared to the type of model presented here, could in principle give
information on the relative importance of radiant and sensible heating.

\subsection{Dynamics of Dirt}

As the snow surface ablates, the dirt layer on it moves (Figure
\ref{dirt}).  We assume the particles are sufficiently small that
 the snow moves purely normal to the
surface. The sideways ($x$-direction) velocity of a  piece of
dirt  is
\begin{equation}
v =-\dot{h} h'
\end{equation}
The thickness of the dirt $s (x) $ must obey a conservation equation
$\dot{s} + \nabla \cdot (vs) = 0 $, since we assume dirt is neither
deposited on nor removed from the surface. The evolution equation for
the thickness of dirt is thus
\begin{equation}
\frac{\partial s}{\partial t} = -\frac{\partial }{\partial x}  (vs) = (\dot{h} h' s)'
\end{equation}
When the surface of the snow is flat ($h' = 0 $) 
the velocity of the dirt $v = 0 $.  Thus the tops of peaks and the
bottoms of valleys are equilibrium points.  The peaks are stable
equilibria where dirt becomes concentrated, while valleys are
unstable (Figure \ref{dirt}).

\subsection{Model}

We now rewrite the model equations incorporating dirt.  
We have equations for the height of the surface
$h $, the dirt thickness $s $,
and the incident power $P $.
\begin{eqnarray}
-\dot{h} & = &\frac{P(x)}{ L } \frac{1} {1 + \frac{C} { \kappa L }
	Ps} +Dh''	 \label{dimheight} \\
\dot{s} & = &  \left (\dot{h} h' s \right)'  
	\label{dimdirt} 
\end{eqnarray}
The only sources of heat flux $P$ we will consider are direct and
reflected radiation.
\begin{equation}
\frac{P (x)} { L } = (1- \alpha e^{-s/s_e})  \frac{I}{ L } 
	+\frac{ \alpha e^{-s/s_e} I}{ \pi L }
	\int_{  {\tiny \begin{array}c \mbox{ line of} \\ \mbox{ sight} 
	\end{array} }  }\frac{dx_1\  (\Delta h - h'(x) \Delta x)  }{ \Delta x ^{ 2} 
	+ \Delta h^{ 2}}
\label{dimpower} 
\end{equation} 

We use the same reference ablation rate as in Section \ref{modelclean}:
$I/L = 10^{-3}\ \mbox{cm}
\ \mbox{sec}^{-1}$. However, the presence of dirt introduces a new
length scale in the problem: the length scale for light absorption by
the dirt. We choose to nondimensionalize in terms of this length,
since the physically important regimes of thin and thick dirt are
measured relative to this thickness. When Driedger measured diameters
of ash particles on a glacier, 90 percent of the particles had
diameters between 0.25 and 1.0 mm\cite{dri81}. We therefore choose $ s_e = 1 $
mm as the order of magnitude extinction length for dirt
absorption; this choice is supported by the good fit to the data. 

The dimensionless timescale comes from combining the ablation rate and
length scale: $\bar t_d = L s_e / I = 100 $ seconds. This is the time
for a depth $s_e $ of snow to melt in bright sun. Glacial debris has $\kappa
\approx 2 \times 10^{ 4}$ erg $\mbox{cm}^{-1} \ \mbox{sec}^{-1}
\:^{\circ}$K \cite{dirtnote}. This allows us to estimate the
dimensionless parameter $\gamma = s_e C I/( L \kappa) = 0.03$.  Note
that the thermal conductivity and the specific heat depend on the
density, wetness, etc. The fit to Driedger's data (Figure
\ref{dirtmelt}) gives a value of $\gamma \approx 0.047$, somewhat
larger than this estimate. We interpret this as a measurement of the
dirt thermal conductivity $\kappa$, and therefore use the implied
value $\kappa=1.3 \times10^{ 4} $ erg $\mbox{cm}^{-1} \mbox{sec}^{-1}
\:^{\circ}$K.  The nondimensionalized diffusion constant is $D \bar t_d /
s_e ^{ 2} = 0.25 $.

For sublimation the time scale $\bar t_d \approx 700$ seconds and the
dimensionless diffusion constant $D \bar t_d /s_e^{2} \approx 1.75$; the
dimensionless parameter $\gamma$ similarly decreases by a factor of 7.

The nondimensionalized equations are
\begin{eqnarray}
\dot{h} & = & -\frac{P}{ 1  +  \gamma Ps} + D \nabla^2 h	 \label{height} \\
\dot{s} & = &  \left ( \dot{h} h' s \right)'  
	\label{dirtmove}\\
P & = & r (1- \alpha e^{-s})  +\frac{ \alpha  e^{-s} r}{ \pi}
 	\int_{  {\tiny \begin{array}c \mbox{ line of} \\ 
	\mbox{ sight} \end{array} }  } \frac{dx_1\  (\Delta h - h'(x) \Delta x) }
	{ \Delta x ^{ 2} + \Delta h^{ 2}}
	\label{power}
\end{eqnarray}
The dimensionless control parameters are $r$, the solar light
intensity; and $s$, the initial dirt thickness. Here we have
introduced the parameter $r$:
\begin{equation}
r = \frac{I}{L} \ 10^{3} \ \mbox{sec/cm}
\end{equation}
to examine the effects of varying the light intensity away from the
typical value.

\subsection{Linear Analysis}
\label{lindirt}

Here we analyze the stability of equations
(\ref{height}--\ref{power}), including effects of dirt. There are two
important regimes: when the initial dirt thickness is small compared
to $s_e$, the dirt acts to modify the reflection-driven
instability. We find that the instability is suppressed by the
absorption of the dirt layer, exponentially in the dirt thickness.  In
this regime, dirt can also a induce travelling, dispersive instability of the
snow surface. Qualitatively, this dispersion arises from the coupling
of dirt motion to absorption. Dirt migrates to the highest point on
the surface---but then the thicker dirt increases the ablation of that
peak, and it ablates until it is no longer a local maximum.  The
existence of these waves is an experimentally testable prediction
which has not, to my knowledge, been discussed before.

The other limit is when the dirt thickness is large compared to
$s_e$. The effective albedo $\alpha e^{-s} \rightarrow 0$. Therefore
the dirt instability is independent of light reflections; the
``light'' therefore acts simply as a source of heat. The instability
is driven by dirt insulating the snow. The characteristic length and
time scale of the instability depends only on the thermal properties
of the dirt. Within this insulation-dominated regime, the behavior of
the instability depends on whether $s \ll 1/(\gamma r)$ or $s \gg
1/(\gamma r)$---see below. Thus there are three different regimes of
behavior, depending the dirt thickness.

As mentioned above, under different weather conditions
uniform heating from the air may be more important than radiant
heating. In this case any amount of dirt slows ablation of 
the snow \cite{rho87}, and the insulation-driven instability is the
only one possible. This can be included in the model by removing the
dirt-dependent absorption of light.

We will perform a linear
perturbation analysis: we assume that {\sl variations} of the dirt
thickness $\Delta s$ are always small. However, the initial uniform
dirt thickness $s_o$ may be large or small relative to $s_e$; this
initial thickness determines the limit of instability.

\subsection{Thin Dirt Limit}

\begin{figure}[!t]
\centerline{\epsfysize=0.8in\epsfbox{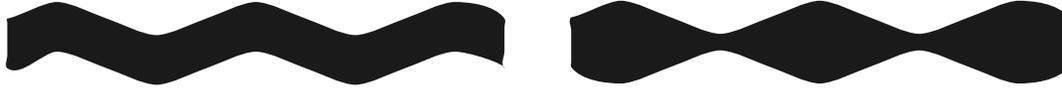}}
\caption[Two Modes of Dirt Modulation]
{ The symmetric (left) and antisymmetric (right) modes of dirt
modulation. The antisymmetric mode is the physically important one
because the symmetric mode is unstable.  }
\label{dirtmode}
\end{figure}

Here we consider the limit $s_o \ll s_e$, meaning the initial uniform
dirt thickness is small compared to the extinction length.

There are in general two modes of dirt modulation (Figure
\ref{dirtmode}): the symmetric mode with constant thickness and the
antisymmetric mode with $\Delta s = 2 \epsilon \cos qx $. The
symmetric mode, because it has constant thickness, it simpler to
analyze. Note that constant dirt
thickness is unstable: any modulation in the dirt thickness
tends to grow.

\subsubsection{Symmetric Mode}

Because the symmetric mode has constant thickness, it insulates the
snow surface uniformly. Therefore, no thick dirt instability can arise
from the symmetric mode. But the symmetric mode affects the
reflection-driven instability.  We look for solutions of the form
\begin{equation}
h  = -mt + \delta e^{\omega t} \cos qx 
\end{equation}
where $m(s)$ is the ablation rate of a flat surface covered with dirt, 
calculated above. The dirt thickness $s_o=$ constant.   We expand the equations
to first order in $\delta$. The resulting dispersion relation is
\begin{equation}
\omega  = \frac {\alpha re^{-s_o} q} {\pi (1+(1-\alpha) \gamma r s_o)^2
	}-Dq^{ 2}. 
\label{dispersion-symmetric}
\end{equation}
Compare this to
the clean snow dispersion relation, Equation \ref {quasilin}. The
first term (proportional to $q $) contains the factor $e^{-s_o} $. This
term decreases exponentially with increasing dirt thickness. For $s_o $
much larger than one, this term is so small that the instability
practically does not exist. The factor $(1+(1-\alpha) \gamma r s_o)^2$ in the
dispersion relation results from uniform insulation by the dirt layer.

\begin{figure}[!t]
\centerline{\epsfysize=3.0in\epsfbox{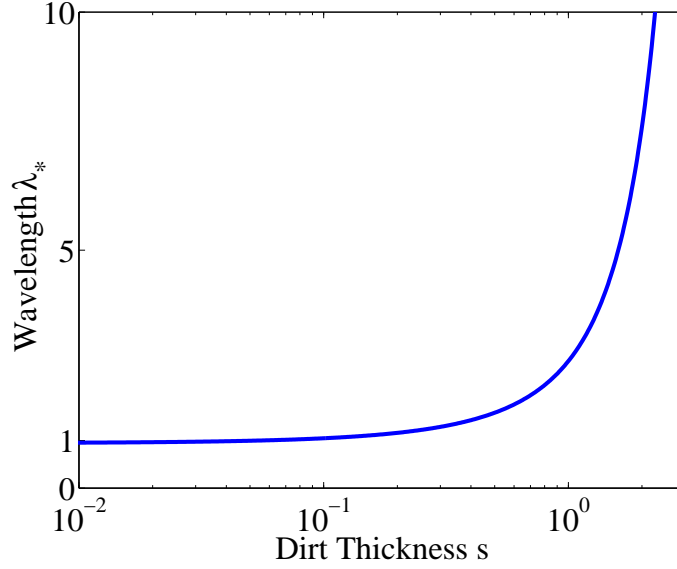}}
\caption[Instability Wavelength with Thin Dirt]
{Wavelength of the Reflection-Driven Instability. The wavelength is
normalized to the most unstable wavelength of clean snow. When $s_o\ll1$, the
wavelength is close to the wavelength in the absence of dirt. However,
the absorption of light by dirt becomes important for $s_o>0.1$ and the
wavelength increases rapidly. The plot is for fixed solar intensity
$r=1$, a typical value.  Parameter values for this plot are as
discussed in the text: $\alpha=0.5$, $D=0.25$, $\gamma=0.047$. }
\label{dirtwave}
\end{figure}

The most unstable mode $q_{ *}$ is
\begin{eqnarray}
q_{ *} & = &\frac{\alpha re^{-s_o} }{2\pi D (1+(1-\alpha) \gamma r s_o)^{ 2}}\\
\omega _{ *} & = & q_{ *}^{ 2} D 
\end{eqnarray}
Figure \ref{dirtwave} shows how dirt cuts off the instability, with
fixed light intensity $r=1$. When $s_o\ll1$, the wavelength is close to
the wavelength in the absence of dirt. However, the absorption of
light by dirt becomes important for $s_o>0.1$ and the wavelength
increases exponentially. As the wavelength increases, the growth rate of the 
instability decreases, and the instability becomes less readily observed.

\subsubsection{Antisymmetric Mode}

\begin{figure}[!t]
\centerline{\epsfysize=3.0in\epsfbox{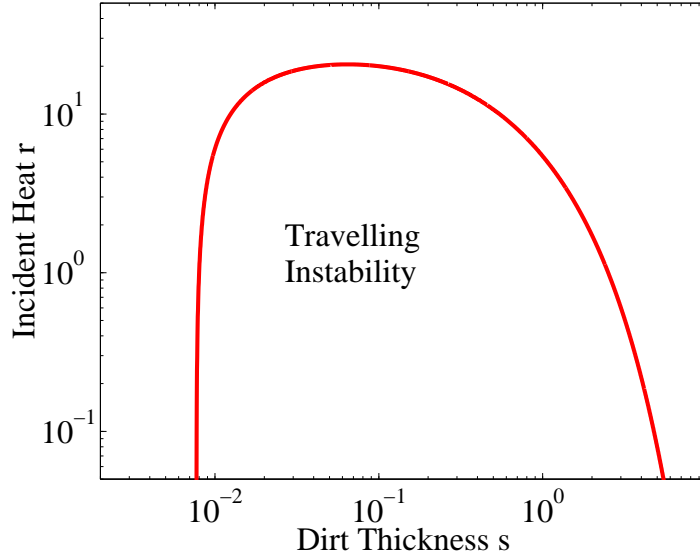}}
\caption[Instability with Thin Dirt]
{Travelling Instability in the Reflection-Driven Instability. Under the
 line, there is an imaginary part of $\omega$, showing the
regime where travelling waves exist. The dirt thickness is normalized
so $s_o=1$ corresponds to one extinction length; similarly, $r=1$ is a
typical intensity of sunlight.  For the typical solar brightness
$r=1$, any dirt thickness $s_o>0.008$ will show a travelling instability;
therefore most dirty snow surfaces should show this
behavior. Parameter values for this plot are as discussed in the text:
$\alpha=0.5$, $D=0.25$, $\gamma=0.047$. }
\label{phase1}
\end{figure}

The antisymmetric mode involves variations in the thickness of the dirt. We
must solve for the coupling between snow ablation and dirt motion. The
solution is of the form
\begin{eqnarray}
h & = &-mt + \delta e^{\omega t} \cos qx \\
s & = & s_o + 2 \epsilon  e^{\omega t} \cos qx 
\end{eqnarray}
where $s$ is the uniform dirt thickness at $t=0$. Upon linearization,
Equation \ref{dirtmove} for the motion of dirt relates the
perturbation amplitudes
\begin{equation}
\frac{ \epsilon }{ \delta } =\frac{ms_oq^{ 2}}{ 2 \omega}.
\end{equation}
The dispersion relation, to second order in $q$, is
\begin{equation}
\omega=[1\pm \sqrt{f}]  \frac{\alpha r e^{-s_o}q}{2 \pi(1+\gamma r' s_o)^2}  - 
	[1 \mp \frac{1}{\sqrt{f}}] \frac{D q^2}{2}
\end{equation}
where $f$ is, defining $w=1/(1+(1-\alpha) \gamma r s_o)$ and recalling
$m=(1-\alpha e^{-s_o}) r w$ is the dimensionless melting rate as a
function of dirt thickness,
\begin{equation}
f= (\alpha r e^{-s_o} \pi^{-1} w^2)^2 (1+ \frac{4 s_o m(m^2 \gamma - \alpha r e^{-s_o}w^2)}{(\alpha r e^{-s_o} \pi^{-1} w^2)^2 }
\end{equation}
In the limit $s_o \rightarrow 0 $, this dispersion relation is identical
to the symmetric mode. However, for increasing dirt thickness it
contains effects from the dirt modulation.  The term $f$ can be {\sl
negative}, leading to an oscillatory  component to
$\omega$. Thus dirt can cause the instability to travel on the snow
surface, in a region of phase space shown in Figure
\ref{phase1}. For the typical solar brightness $r=1$, any dirt
thickness $s_o>0.008$ will induce travelling; therefore, most dirty snow
surfaces should show this behavior.  Qualitatively, this arises from
the coupling of dirt motion to absorption. Dirt migrates to the
highest point on the surface---but then the thicker dirt increases the
ablation of that peak, and it ablates until it is no longer a local
maximum. The positive and negative roots in the dispersion relation
correspond to left and right moving modes. The existence of these
travelling instabilities is an experimentally testable prediction.

Note that the equation is not well-behaved for $f=0$. When $f=0$ the
terms in the equation coupling motion of dirt to ablation vanish; the
dispersion relation reduces to the expression for the symmetric mode above.

When $f$ is negative, we can  find the fastest
growing wavelength by looking at the real part of
$\omega$:
\begin{eqnarray}
q_* &=& \frac{\alpha e^{-s_o} r }{2 \pi D (1+(1-\alpha) \gamma r s_o)^2}	\\
\omega_* &=& \frac{D}{2} q_*^2
\end{eqnarray}

\subsection{Thick Dirt Limit}

The equations are considerably simplified in the limit of thick dirt
$s_o\gg1$. The effective albedo $\alpha e^{-s_o} \rightarrow 0$. Therefore
the dirt instability is independent of any reflections; the quasi-linearized
equations are truly linear in this limit. The thick-dirt instability
is driven purely by dirt motion coupled to slower ablation under a
thicker dirt layer. This instability is the linear precursor to the
dirt cones of Figure \ref{dirtpic}.

Note that if light is not an important source of heat, the ``thick
dirt limit'' is actually valid for all dirt thicknesses.

\begin{figure}[!ht]
\centerline{\epsfysize=3.0in\epsfbox{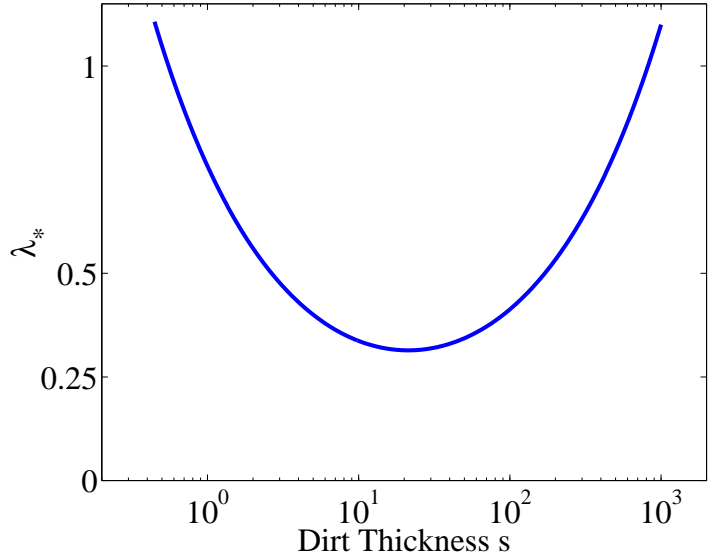}}
\caption[Most Unstable Wavelength vs. Dirt Thickness]
{ Most unstable wavelength $\lambda_*$ versus dirt thickness $s_o$,
with typical heat flux $r=1$. The wavelength is normalized to the most
unstable wavelength of clean snow. Comparing this figure to the
thin-dirt instability, we see that when $s_o>1$ the wavelength will
initially decrease, then increase beyond $s_o=20$. The growth rate of
this instability will be greatest where the wavelength is smallest.  }
\label{lamplot2}
\end{figure}

Replacing $\alpha e^{-s_o} \rightarrow 0$, the symmetric mode
disappears. The background ablation rate $m= r/(1+\gamma r s_o)$.  The
dispersion relation is, to second order in $q$,
\begin{equation}
\omega = \pm \frac{\sqrt{\gamma s_o m^3}}{2} q - \frac{D}{2}
	 q^2
\end{equation}
Here no imaginary component to the dispersion relation is present; it
is a straightforward linear instability with one growing mode. The
most unstable wavenumber is
\begin{equation}
q_*= \frac{1}{2D} \sqrt{\frac{\gamma r^3 s_o}{(1+\gamma r s_o)^3}} 
	\end{equation}
For a fixed value of the heat input $r$, the most unstable wavelength
scales differently at small and large $s_o$:
\begin{eqnarray}
\lambda_* &\sim& s_o^{-1/2} \ \mbox{for} \ s_o \ll 1/(\gamma r) \\
&\sim& s_o \ \mbox{for} \ s_o \gg 1/(\gamma r)
\end{eqnarray}
The location of the minimum wavelength is determined by the
dimensionless parameter $\gamma$, which represents how well the snow
insulates per unit thickness. Therefore, even for optically thick dirt
$s_o\gg1$ there is a change in the behavior, depending on the value of
$s_o$ compared to the insulation parameter.  
Since typically $\gamma r
= 0.05$, these limits are consistent.

There is an optimal $s_o \sim 1/(\gamma r) \approx 2$ cm where the
wavelength is smallest. Figure \ref{lamplot2} illustrates this: it
shows the unstable wavelength vs. dirt thickness for the typical
$r=1$, with the optimal $s_o \approx 20 \approx 2$ cm. Comparing
this figure to the thin-dirt instability, we see that when $s_o>1$ the
wavelength will initially decrease, then increase beyond $s_o=20$. The
growth rate of this instability will be greatest where the wavelength
is smallest.

\section{Discussion: Comparison to Experiment}

This paper has presented work on a simple theory to describe the
initial formation ablation structures such as suncups, penitentes and
dirt cones. We have tried to make the model as simple as possible
while including the essential physics. As we have shown, most
parameters in the equations can be calculated or measured in
experiments, allowing predictions with no free parameters. The
exception is the effective diffusion coefficient $D$, which we
estimate using the value for light diffusion. However, we have not
realistically treated the small-scale scattering of light in these
schematic results.

At this point, the only quantitative comparison between this model and
experiment is the prediction of ablation rate of a flat snow surface,
compared with the data of Driedger in Figure \ref {dirtmelt}. This
measurement allows us to extract the dimensionless constant governing
dirt insulation. The good agreement indicates we have captured the
important effects of dirt.

The linear stability analysis of the equations shows the two types of
instability described in the literature. The model predicts the
dependence of the most unstable wavelength and characteristic growth
rate on the experimental control parameters, predictions which could
be tested. We argue that for little or no surface
dirt, light reflection drives the instability. This instability
is exponentially suppressed by a dirt layer, consistent with field
observations. We predict travelling modes induced by a modulated dirt
layer in this regime. The existence of such travelling modes is an
experimentally testable new phenomenon.

In the presence of a thick layer of dirt, our analysis finds the
insulation-driven instability, as expected. Here we predict an optimal 
dirt thickness where the instability is most easily observed, which
depends on the thermal properties of the dirt.

The visually striking structures in the field are the larger
structures: penitentes and dirt cones. Understanding the nonlinear
regime of the model presented here is therefore of interest, and will
be the subject of a future paper. The scale of both penitentes and
dirt cones is typically larger than the size of smaller-amplitude
structures. One way to explain this, which has been suggested from
observations\cite{lli54,wil53}, is that large structures grow at the
expense of small ones. Such coarsening behavior is also apparent in
preliminary work on the nonlinear regime of the model presented here.

The most obvious problem with the results here is that we have
considered variation of the surface height in only one
direction. Checking whether the results are the same for a realistic
2D surface is a necessary extension of this work.
A better understanding of the small-scale cutoff is also
important. In particular, we need to understand how using different
representations of the short-scale physics affect the numerical
predictions (of the fastest-growing wavelength, for example). 

Because the model here is simplified, we have left out some physical
effects which may be important in the experiment.  Our treatment of
light reflection considered single reflections only, which may be a
bad approximation with the albedo is close to 1 (large amount
reflected). In the field, the sun of course is not always high
overhead---the variation of the angle of incident light over the
course of the day might change the shapes. Other possibly important
effects which can occur in field situations include other sources of
heat transfer to the surface, gravity, and the deposition/removal of
dirt. Better comparison with lab or field experiments should indicate
which of these effects are most important to include.
 
Acknowledgements: I am grateful to John Wettlaufer and Norbert
Untersteiner for comments on this paper. I thank Eric Nodwell and Tom
Tiedje for discussing their work on this problem with me. I also wish
to thank Michael Brenner, Daniel Fisher, David Weitz, Martine Benamar,
David Lubensky, and David Nelson for helpful discussions, questions,
and criticism.  This work was supported by the NSF under grant
DMS9733030.


\begin{thebibliography}{10}
\bibitem{langbarr}
The articles describing these experiments are in Japanese, which the author
  unfortunately does not understand.
\bibitem{theorynote}
Rhodes, Armstrong, and Warren \cite{rho87} report two ``initial attempts at
  modeling'' published in German. Currently Nodwell and Tiedje are
  investigating the formation of suncups (personal communication). They
  consider in detail the scattering of light in the snowpack, and deduce a
  range of allowable suncup sizes from this.
\bibitem{lightnote}
Note that the 1 cm extinction length comes primarily from near-infrared
  wavelengths---the wavelengths of light which make the dominant contribution
  to the albedo and to ablation. Shorter wavelengths of visible light can
  penetrate much farther into the snow, 20 to 50 cm in old snow. See Wiscombe
  and Warren \cite{wis80} for a thorough discussion of these effects.
\bibitem{heatdiff}
One might think that diffusion of heat through the snow can be an important
  stabilizing factor. However, a simple calculation shows that this is not
  sufficient to provide a short-wavelength cutoff. I coupled temperature
  diffusion in the snow with heat conservation on the interface, assuming
  constant temperature in the air. This gives a dispersion relation with a $q$
  term from reflections, as in the text, and a term proportional to
  $-\sqrt{1+q^2}$ from diffusion in the snow. As a result of this $q$
  dependence, at short wavelengths the stabilization by heat diffusion is
  controlled by a dimensionless parameter $\Delta = (T_s-T_{\infty})C\ L^{-1}$,
  where $T_s-T_{\infty}$ is the temperature difference between the surface of
  the snow and deep in the snow, $C$ is the specific heat, and $L$ the latent
  heat. In order for heat diffusion in the snow to stabilize short wavelengths,
  we must have $\Delta \geq 0.47$. However, $C/L \approx 6 \times 10^{-3}
  /\:^{\circ}$K. Thus for a temperature difference of 10 $\:^{\circ}$C the
  parameter $\Delta \approx 6 \times 10^{-2}$, which is far too low to provide
  the cutoff.
\bibitem{meltnote}
If most melting takes place during four hours of the day when the sun is most
  intense, this number gives 14 cm/day for the melting rate of the snow. It is
  useful to compare this to the field measurements of Kotlyakov and Lebedeva on
  penitente formation \cite{kot74}. If we divide their measurement of the total
  radiation received in one day by the solar intensity at noon, we would get
  3.5 hours estimated illumination time. They typically found snow height
  decreases of 6 cm/day. Thus our estimate may be too high, but is the correct
  order of magnitude. Kotlyakov and Lebedeva found 1.5 m high penitentes formed
  in 24 days.
\bibitem{extnote}
When defined this way, $s_e $ is an extinction length for reflections, which
  differs by a factor of two from the transmission extinction length.
\bibitem{dirtnote}
Drewry \cite{dre72} deduces a value of 5.6 $\times\ 10^{ 4}$ erg
$\mbox{cm}^{-1} \ \mbox{sec}^{-1} \:^{\circ}$K for debris covering
dirt cones, which he compares to other measurements for sand around 2
$\times\ 10^{ 4}$ erg $\mbox{cm}^{-1} \ \mbox{sec}^{-1} \:^{\circ}$K.
\end{thebibliography}

\end{document}